\documentstyle[epsfig,a4p,12pt,rotating]{article}
\begin{document}
%
%
\newcommand{\PPEnum}    {CERN-EP/98-093}
\newcommand{\PNnum}     {OPAL Physics Note PN-xxx}
\newcommand{\TNnum}     {OPAL Technical Note TN-xxx}
\newcommand{\Date}      {16 June 1998}
\newcommand{\Author}    {M.~J.~Oreglia, A.~S.~Turcot}
\newcommand{\MailAddr}  {Andre.Turcot@cern.ch}
\newcommand{\EdBoard}   {K. Bell, K. Desch, P. Maettig, G. Wilson}
\newcommand{\DraftVer}  {FINAL DISPATCH DRAFT}
\newcommand{\DraftDate} {\Date}
\newcommand{\TimeLimit} {Monday, 8 June 1998, 17h00 Geneva time}

\def\toprule{\noalign{\hrule \medskip}}
\def\midrule{\noalign{\medskip\hrule }}
\def\botrule{\noalign{\medskip\hrule }}
\setlength{\parskip}{\medskipamount}

\newcommand{\TPONE}{$\theta_{\gamma 1}$}
\newcommand{\TPTWO}{$\theta_{\gamma 2}$}
\newcommand{\CTPM}{$\cos\theta_{\mrm miss}$}
\newcommand{\ECV} {$E_{\mrm{excess}}$}
\newcommand{\NPHO} {$\mrm N_{\gamma} > 1$}

\newcommand{\mgg} {$M_{\gamma \gamma}$}
\newcommand{\mdip} {M_{\gamma \gamma}}
\newcommand{\Stogg} {${\mathrm S} \ra \gamma \gamma$}
\newcommand{\epem}{{\mathrm e}^+ {\mathrm e}^-}
\newcommand{\tptm}{{\tau}^+ {\tau}^-}
\newcommand{\mpmm}{{\mu}^+ {\mu}^-}

\newcommand{\qqbar}{{\mathrm q}\bar{\mathrm q}}
\newcommand{\nn}{\nu \bar{\nu}}
\newcommand{\nunu}{\nu \bar{\nu}}
\newcommand{\mumu}{\mu^+ \mu^-}
\newcommand{\ellell}{\ell^+ \ell^-}
\newcommand{\MZ}{M_{\mathrm Z}}
\newcommand{\MH}{M_{\mathrm H}}
\newcommand{\MX} {M_{\mathrm{X}}}
\newcommand{\MY} {M_{\mathrm{Y}}}

\newcommand {\Hboson}        {{\mathrm H}^{0}}
\newcommand {\Hzero}         {${\mathrm H}^{0}$}
\newcommand {\Zboson}        {{\mathrm Z}^{0}}
\newcommand {\Zzero}         {${\mathrm Z}^{0}$}
\newcommand {\Wpm}           {{\mathrm W}^{\pm}}
\newcommand {\allqq}         {\sum_{q \neq t} q \bar{q}}
\newcommand {\mixang}        {\theta _{\mathrm {mix}}}
\newcommand {\thacop}        {\theta _{\mathrm {Acop}}}
\newcommand {\cosjet}        {\cos\thejet}
\newcommand {\costhr}        {\cos\thethr}
\newcommand{\epair}    {\mbox{${\mathrm e}^+{\mathrm e}^-$}}
\newcommand{\mupair}   {\mbox{$\mu^+\mu^-$}}
\newcommand{\taupair}  {\mbox{$\tau^+\tau^-$}}
\newcommand{\qpair}    {\mbox{${\mathrm q}\overline{\mathrm q}$}}
\newcommand{\ff}       {{\mathrm f} \bar{\mathrm f}}
\newcommand{\gaga}     {\gamma\gamma}
\newcommand{\WW}       {{\mathrm W}^+{\mathrm W}^-}
\newcommand{\eeee}     {\mbox{\epair\epair}}
\newcommand{\eemumu}   {\mbox{\epair\mupair}}
\newcommand{\eetautau} {\mbox{\epair\taupair}}
\newcommand{\eeqq}     {\mbox{\epair\qpair}}
\newcommand{\eeff}     {\mbox{$\mathrm e^+e^- f \bar{\mathrm f}$}}
\newcommand{\llnunu}   {\mbox{\lpair\nu\nubar}}
\newcommand{\lnuqq}    {\mbox{\lept\nubar\qpair}}
\newcommand{\zee}      {\mbox{Zee}}
\newcommand{\wenu}     {\mbox{We$\nu$}}

\newcommand{\el}       {\mbox{${\mathrm e}^-$}}
\newcommand{\selem}    {\mbox{$\tilde{\mathrm e}^-$}}
\newcommand{\smum}     {\mbox{$\tilde\mu^-$}}
\newcommand{\staum}    {\mbox{$\tilde\tau^-$}}
\newcommand{\slept}    {\mbox{$\tilde{\ell}^\pm$}}
\newcommand{\sleptm}   {\mbox{$\tilde{\ell}^-$}}
%
%
\newcommand{\zz}        {\mbox{$|z_0|$}}
\newcommand{\dz}        {\mbox{$|d_0|$}}
\newcommand{\sint}      {\mbox{$\sin\theta$}}
\newcommand{\cost}      {\mbox{$\cos\theta$}}
\newcommand{\mcost}     {\mbox{$|\cos\theta|$}}
\newcommand{\dedx}      {\mbox{$dE/dx$}}
\newcommand{\wdedx}     {\mbox{$W_{dE/dx}$}}
\newcommand{\xe}        {\mbox{$x_{\rm E}$}}
\newcommand{\xgam}      {x_{\gamma}}
\newcommand{\Mrec}      {M_{\mrm{recoil}}}
\newcommand{\Lgg}       {\mbox{${\cal L}(\gaga)$}}
\newcommand{\ggjj}{\mbox{$\gaga\;{\rm jet-jet}$ }}
\newcommand{\ggqq}{\mbox{$\gaga {\rm q\bar{q}}$ }}
\newcommand{\ggll}{\mbox{$\gaga\ell^{+}\ell^{-}$}}
\newcommand{\ggnn}{\mbox{$\gaga\nu\bar{\nu}$}}
\newcommand{\eeeell}{\mbox{\epair$\rightarrow$\epair\lpair}}
\newcommand{\eell}{\mbox{\epair\lpair}}
\newcommand{\llgam}{\mbox{$\ell\ell(\gamma)$}}
\newcommand{\nunugam}{\mbox{$\nu\bar{\nu}\gaga$}}
\newcommand{\acope}{\mbox{$\Delta\phi_{\mathrm{EE}}$}}
\newcommand{\nee}{\mbox{N$_{\mathrm{EE}}$}}
\newcommand{\eesum}{\mbox{$\Sigma_{\mathrm{EE}}$}}
\newcommand{\acoph}{\mbox{$\Delta\phi_{\mathrm{HCAL}}$}}
\newcommand {\mm}         {\mu^+ \mu^-}
\newcommand {\emu}        {\mathrm{e}^{\pm} \mu^{\mp}}
\newcommand {\et}         {\mathrm{e}^{\pm} \tau^{\mp}}
\newcommand {\mt}         {\mu^{\pm} \tau^{\mp}}
\newcommand {\lemu}       {\ell=\mathrm{e},\mu}
\newcommand {\Zz}         {\mbox{${\mathrm{Z}^0}$}}
%
\newcommand{\Ecm}         {\mbox{$E_{\mathrm{cm}}$}}
\newcommand{\Ebeam}       {E_{\mathrm{beam}}} 
\newcommand{\ipb}         {\mbox{pb$^{-1}$}}

\newcommand{\gsim}{\;\raisebox{-0.9ex}
           {$\textstyle\stackrel{\textstyle >}{\sim}$}\;}
\newcommand{\lsim}{\;\raisebox{-0.9ex}{$\textstyle\stackrel{\textstyle<}
           {\sim}$}\;}

\newcommand{\degree}    {^\circ}
%
\newcommand{\roots}     {\sqrt{s}}
%
%
\newcommand{\thmiss}    { \theta_{miss} }
\newcommand{\cosmiss}   {| \cos \thmiss |}
%
%
\newcommand{\Evis}      {\mbox{$E_{\mathrm{vis}}$}}
\newcommand{\Rvis}      {\mbox{$R_{\mathrm{vis}}$}}
\newcommand{\Mvis}      {\mbox{$M_{\mathrm{vis}}$}}
\newcommand{\Rbal}      {\mbox{$R_{\mathrm{bal}}$}}
%
%
%
\newcommand{\PhysLett}  {Phys.~Lett.}
\newcommand{\PRL}       {Phys.~Rev.\ Lett.}
\newcommand{\PhysRep}   {Phys.~Rep.}
\newcommand{\PhysRev}   {Phys.~Rev.}
\newcommand{\NPhys}     {Nucl.~Phys.}
\def\NIM                {\mbox{Nucl. Instrum. Meth.}}
\newcommand{\NIMA}[1]   {\NIM\ {\bf A{#1}}}
\newcommand{\IEEENS}    {IEEE Trans.\ Nucl.~Sci.}
\newcommand{\ZPhysC}[1]    {Z. Phys. {\bf C#1}}
\newcommand{\EurPhysC}[1]    {Eur. Phys. {\bf C#1}}
\newcommand{\PhysLettB}[1] {Phys. Lett. {\bf B#1}}
\newcommand{\CPC}[1]      {Comp.\ Phys.\ Comm.\ {\bf #1}}
\def\etal{\mbox{{\it et al.}}}
%
%
\newcommand{\OPALColl}  {OPAL Collab.}
\newcommand{\JADEColl}  {JADE Collab.}
%
\newcommand{\onecol}[2] {\multicolumn{1}{#1}{#2}}
\newcommand{\colcen}[1] {\multicolumn{1}{|c|}{#1}}
\newcommand{\ra}        {\rightarrow}   
\newcommand{\ov}        {\overline}   
\def\mrm       {\mathrm}
\newcommand {\downto}
        {\mbox{ \begin{picture}(14,10)
                   \put(0,10){\line(0,-1){5.0}}
                   \put(2,5){\oval(4,4)[bl]}
                   \put(1,0){\makebox(0,0)[bl]{$\rightarrow$}}
                \end{picture} }}


\begin{titlepage}
\begin{center}{\large   EUROPEAN LABORATORY FOR PARTICLE PHYSICS
}\end{center}\bigskip
\begin{flushright}
       \PPEnum   \\ \DraftDate
\end{flushright}
\bigskip\bigskip\bigskip\bigskip\bigskip
%
%
\begin{center}{\huge\bf\boldmath Search for Higgs Bosons and New \\
                        Particles Decaying into Two Photons \\
\vspace{3pt}
                               at $\sqrt{s} = 183$ GeV
}\end{center}\bigskip\bigskip
\begin{center}{\LARGE The OPAL Collaboration
}\end{center}\bigskip\bigskip
%
%
\bigskip\begin{center}{\large  Abstract}\end{center}
A search for the resonant production of high mass photon pairs
associated with a leptonic or hadronic system has been performed using 
a data sample of 57.7~\ipb\ collected at an average center-of-mass energy
of 182.6 GeV with the OPAL detector at LEP. No evidence for contributions
from non-Standard Model physics processes was observed.
The observed candidates are used to place limits on $B$($\Hboson \ra \gaga$) 
assuming a Standard Model production rate for Higgs boson masses up to 92 GeV,
and on the production cross section for a scalar resonance decaying
into di-photons up to a mass of 170 GeV. Upper limits on the
product of cross section and branching ratios,
$\sigma(\epem\ra {\mrm X Y})\times B({\mrm X} \ra \gaga)\times 
B(\mrm Y \ra f\bar{f})$,
as low as 70~fb are obtained over the range $10 < \MX < 170$~GeV for the case
where $10 < \MY < 160$ GeV and $\MX+\MY>90$~GeV, independent of 
the nature of Y provided it decays to a fermion pair and has
negligible width.
Higgs scalars which couple only to gauge bosons at Standard Model
strength are ruled out up to a mass of 90.0~GeV at the 95\% confidence level.
Limits are also placed on non-minimal Higgs sectors having triplet
representations.



\bigskip\bigskip\bigskip\bigskip

\begin{center}{\large
(Submitted to Physics Letters B)
}\end{center}

\end{titlepage}

\begin{center}{\Large        The OPAL Collaboration
}\end{center}\bigskip
\begin{center}{
K.\thinspace Ackerstaff$^{  8}$,
G.\thinspace Alexander$^{ 23}$,
J.\thinspace Allison$^{ 16}$,
N.\thinspace Altekamp$^{  5}$,
K.J.\thinspace Anderson$^{  9}$,
S.\thinspace Anderson$^{ 12}$,
S.\thinspace Arcelli$^{  2}$,
S.\thinspace Asai$^{ 24}$,
S.F.\thinspace Ashby$^{  1}$,
D.\thinspace Axen$^{ 29}$
% ...
}\end{center}\bigskip
\bigskip
$^{  1}$School of Physics and Astronomy, University of Birmingham,
Birmingham B15 2TT, UK
\newline
$^{  2}$Dipartimento di Fisica dell' Universit\`a di Bologna and INFN,
I-40126 Bologna, Italy
\newline
$^{  3}$Physikalisches Institut, Universit\"at Bonn,
D-53115 Bonn, Germany
\newline
$^{  4}$Department of Physics, University of California,
Riverside CA 92521, USA
\newline
$^{  5}$Cavendish Laboratory, Cambridge CB3 0HE, UK
\newline
$^{  6}$Ottawa-Carleton Institute for Physics,
Department of Physics, Carleton University,
Ottawa, Ontario K1S 5B6, Canada
\newline
$^{  7}$Centre for Research in Particle Physics,
Carleton University, Ottawa, Ontario K1S 5B6, Canada
\newline
$^{  8}$CERN, European Organisation for Particle Physics,
CH-1211 Geneva 23, Switzerland
\newline
$^{  9}$Enrico Fermi Institute and Department of Physics,
University of Chicago, Chicago IL 60637, USA
\newline
$^{ 10}$Fakult\"at f\"ur Physik, Albert Ludwigs Universit\"at,
D-79104 Freiburg, Germany
\newline
$^{ 11}$Physikalisches Institut, Universit\"at
Heidelberg, D-69120 Heidelberg, Germany
\newline
$^{ 12}$Indiana University, Department of Physics,
Swain Hall West 117, Bloomington IN 47405, USA
\newline
$^{ 13}$Queen Mary and Westfield College, University of London,
London E1 4NS, UK
\newline
$^{ 14}$Technische Hochschule Aachen, III Physikalisches Institut,
Sommerfeldstrasse 26-28, D-52056 Aachen, Germany
\newline
$^{ 15}$University College London, London WC1E 6BT, UK
\newline
$^{ 16}$Department of Physics, Schuster Laboratory, The University,
Manchester M13 9PL, UK
\newline
$^{ 17}$Department of Physics, University of Maryland,
College Park, MD 20742, USA
\newline
$^{ 18}$Laboratoire de Physique Nucl\'eaire, Universit\'e de Montr\'eal,
Montr\'eal, Quebec H3C 3J7, Canada
\newline
$^{ 19}$University of Oregon, Department of Physics, Eugene
OR 97403, USA
\newline
$^{ 20}$Rutherford Appleton Laboratory, Chilton,
Didcot, Oxfordshire OX11 0QX, UK
\newline
$^{ 22}$Department of Physics, Technion-Israel Institute of
Technology, Haifa 32000, Israel
\newline
$^{ 23}$Department of Physics and Astronomy, Tel Aviv University,
Tel Aviv 69978, Israel
\newline
$^{ 24}$International Centre for Elementary Particle Physics and
Department of Physics, University of Tokyo, Tokyo 113, and
Kobe University, Kobe 657, Japan
\newline
$^{ 25}$Institute of Physical and Environmental Sciences,
Brunel University, Uxbridge, Middlesex UB8 3PH, UK
\newline
$^{ 26}$Particle Physics Department, Weizmann Institute of Science,
Rehovot 76100, Israel
\newline
$^{ 27}$Universit\"at Hamburg/DESY, II Institut f\"ur Experimental
Physik, Notkestrasse 85, D-22607 Hamburg, Germany
\newline
$^{ 28}$University of Victoria, Department of Physics, P O Box 3055,
Victoria BC V8W 3P6, Canada
\newline
$^{ 29}$University of British Columbia, Department of Physics,
Vancouver BC V6T 1Z1, Canada
\newline
$^{ 30}$University of Alberta,  Department of Physics,
Edmonton AB T6G 2J1, Canada
\newline
$^{ 31}$Duke University, Dept of Physics,
Durham, NC 27708-0305, USA
\newline
$^{ 32}$Research Institute for Particle and Nuclear Physics,
H-1525 Budapest, P O  Box 49, Hungary
\newline
$^{ 33}$Institute of Nuclear Research,
H-4001 Debrecen, P O  Box 51, Hungary
\newline
$^{ 34}$Ludwigs-Maximilians-Universit\"at M\"unchen,
Sektion Physik, Am Coulombwall 1, D-85748 Garching, Germany
\newline
\bigskip\newline
$^{  a}$ and at TRIUMF, Vancouver, Canada V6T 2A3
\newline
$^{  b}$ and Royal Society University Research Fellow
\newline
$^{  c}$ and Institute of Nuclear Research, Debrecen, Hungary
\newline
$^{  d}$ and Department of Experimental Physics, Lajos Kossuth
University, Debrecen, Hungary
\newline
$^{  e}$ on leave of absence from the University of Freiburg
\newline

\newpage
\section{Introduction}
\label{sec:intro}

A search is described for the production of a di-photon
resonance produced in $\epem$ collisions at 
an average center-of-mass energy $\roots$ = 182.6 GeV using an
integrated luminosity of 57.7~\ipb\
taken with the OPAL detector at LEP. 
We consider the process 
$\epem \ra \mrm X Y$, with $\mrm X \ra \gaga, Y \ra \ff $
where $\ff$ may be quarks, charged leptons, or a neutrino pair.
The analysis considers two hypotheses for Y. 
The first  
is where Y is required to be consistent with a \Zzero; in this
case, particle X can be taken to be a Higgs boson. 
The second case is a general
search where Y is assumed to be a particle 
with no requirement imposed its mass;  
the search is therefore sensitive, for example, to the pair 
production of Higgs bosons.

In the case of the Standard Model Higgs boson,
$\Hboson \ra \gaga$ proceeds via a top quark or W boson loop
and the rate is too small for observation at existing accelerators even for a 
kinematically accessible Higgs boson.
A 90 GeV Higgs boson, for example, has an expected di-photon
branching ratio of ${\cal O}(10^{-3})$~\cite{HBR}. 
In the framework of effective Lagrangians, 
anomalous Higgs couplings can be generated by dimension-6 operators
and may result in large production cross section 
and/or di-photon branching ratio~\cite{Hagiwara}. 
For non-minimal Higgs sectors, a large $\Hboson \ra \gaga$ branching
ratio can
arise in a number of scenarios. As discussed in 
reference \cite{Fermiophobic}, these include the ``Bosonic'' Higgs 
model~\cite{Bosonic}, Type I Two-Higgs Doublet 
models with fermiophobic couplings~\cite{TypeI}
and the Higgs Triplet model \cite{Triplets}.
In the Higgs Triplet model (HTM), in addition to the standard 
complex scalar doublet with hypercharge $Y=0$,
there is a complex $Y=2$ triplet and a real $Y=0$ triplet.
The
triplets have equal vacuum expectation values. With these choices, 
custodial $\mrm SU(2)$ symmetry is maintained and the 
experimental constraint on 
$\rho \equiv M_{\mrm W}/ (M_{\mrm Z}\cos\theta_W)$
is therefore satisfied.



The search considered three topologies. The first was a search for a system 
of two photons recoiling from a hadronic system.  The second topology was
a search for di-photons produced in association with a low multiplicity
system of charged tracks consistent with the production of a pair of charged 
leptons.
The third topology was a search for no significant detector activity
other than a pair of photons recoiling from unobserved
neutral particles. For the \Hzero\Zzero\ final state, the 
recoil mass
was required to be consistent with the \Zzero\ mass; the more general
search relaxed this requirement. 

A background common to all topologies was from events with two
visible 
initial state radiation (ISR) photons resulting in an on-shell 
\Zzero\ recoiling from a di-photon system:
\begin{center}
$\epem \ra \Zboson (\gaga)_{\mrm ISR}$, $\Zboson \ra {\mrm f} \bar{\mrm f}$.
\end{center}
For photons visible in the detector, the cross section is the order
of 200-400 fb, depending on the photon kinematics. High mass photon
pairs from ISR (\mgg$>40$ GeV ) comprise roughly 50\% of the visible 
$\Zboson\gaga$ cross 
section; however, they will not exhibit any resonant structure.
 

To assess the sensitivity of the analysis, two production models were
considered: the Standard Model process $\epem \ra \Hboson\Zboson$ 
and Two Higgs Doublet models (2HDM) for $\epem \ra \Hboson \mrm A^0$. 
Throughout 
this paper, ``\Hzero'' refers to a neutral CP-even scalar where 
non-minimal Higgs sector models are discussed. For the case of the HTM 
model, following the notation of reference \cite{Gunion}, \Hzero\ 
represents the fermiophobic neutral Higgses, the $\mrm H_5^0$ and 
$\mrm H^{'0}_1$. 
Similarly, the $\mrm A^0$ 
refers to the CP-odd neutral Higgs or the $\mrm H_3^0$ for the case of the 
HTM.

There are existing limits on the production of a di-photon resonance 
which couples 
to the \Zzero. Using data taken up to $\sqrt{s}=172$ GeV, OPAL has set upper 
limits on the branching ratio  
$\Hboson \ra \gaga$ assuming a Standard Model production rate for
Higgs masses up to 77 GeV \cite{OPAL_HGG}. When interpreted within the
Bosonic Higgs model, the above limits 
correspond to a 95\% confidence level (CL) lower mass limit of 76.5 GeV  for
such a Higgs scalar. This analysis exploits the higher center-of-mass
energy and improvements in the photon detection efficiency.

For the mass interval $10 < \MX < 40$ GeV, the most stringent limits for 
production of a scalar resonance decaying into two photons
are obtained from studies of \Zzero\ decays \cite{OPAL_ggjj_1,LowMgg}.
For a Standard Model production rate, the corresponding upper limits
for $B(\Hboson \ra \gaga)$ are in the range 0.008 to 0.02 at the 95\% CL.
Other searches for the production of a massive di-photon resonance in \Zzero\
decays are described in references \cite{other_gg}.


\section{Data and Monte Carlo Samples}
%
The analysis was performed on the data collected with the OPAL detector 
\cite{detector}
during the 1997 LEP run.
The sample consisted of an integrated luminosity of $57.7$ \ipb\
collected at a luminosity weighted center-of-mass energy equal to 
$182.6 \pm 0.03$ GeV. The data were acquired at 
center-of-mass energies ranging from 181 to 184 GeV.

The HZHA generator \cite{HZHA}
was used to 
simulate the process $\epem \ra \Hboson\Zboson$ followed
by $\Hboson \ra \gaga$ for each \Zzero\ decay channel. 
For the general search, a mass grid was generated using 
the $\epem \ra \Hboson \mrm A^0$ process as a model 
for the $\epem \ra {\mrm XY} \ra \gamma\gamma+\mrm f\bar{f}$ final state.
The grid was generated for X masses from 10 to 170 GeV and Y masses
from 10 to 160 GeV such that $\MX + \MY > M_{\mrm Z}$. This latter
constraint was motivated by the higher sensitivity of searches performed
at $\roots =  M_{\mrm Z}$ for lower masses. The grid was generated
 in 10~GeV steps near the
kinematic boundaries and the sensitivity at intermediate points
was determined by interpolation. 


The Standard Model backgrounds from
$\epem \ra (\gamma/{\rm Z)^{\ast} \ra {\rm q\bar{q}}} $ were simulated
using the PYTHIA \cite{PYTHIA} package with the set of hadronization 
parameters described in reference \cite{jtparams}. As a cross check of
the treatment of initial state radiation, samples were also generated using the 
HERWIG \cite{HERWIG} generator. The programs BHWIDE
\cite{BHWIDE} and TEEGG \cite{TEEGG} were employed
to model the background from Bhabha scattering. 
The processes $\epem \ra \ellell$ 
with $\ell \equiv \mu , \tau$ were simulated using
KORALZ \cite{KORALZ}. The KORALZ program was also used to 
generate events of the type $\epem\ra\nu\ov{\nu}\gamma(\gamma)$.
The process $\epem \ra \gaga$ was simulated using the
RADCOR generator~\cite{RADCOR}.
Purely leptonic four-fermion processes of the type  
$\epem \ell^+ \ell^-$, where $\ell \equiv {\rm e},\mu,\tau$, were
modelled using the Vermaseren \cite{VERMASEREN} and grc4f \cite{grc4f} 
generators. Other four-fermion processes were modelled using the 
grc4f and 
EXCALIBUR \cite{excalibur} event generators.
The $\mrm W^{\pm}e^{\mp} \nu$
final state was modelled using the KORALW generator \cite{KORALW}.
Both signal and background events were processed using the full
OPAL detector simulation~\cite{GOPAL} and analyzed in the 
same manner as the data.



\section{Event Selection}

For all topologies, charged tracks (CT) and unassociated 
electromagnetic calorimeter (EC) clusters were 
defined as those satisfying the criteria defined in reference \cite{CTSEL},
unless otherwise specified. For each channel, preselection cuts were 
applied which employed the following measured quantities:
\begin{itemize}
\item $\Evis$ and $\vec{p}_{\mrm{vis}}$: the scalar and vector sums
      of charged track momenta, unassociated EC and hadron calorimeter cluster
      energies.
\item $\Rvis \equiv \frac{\mbox{\Evis}}{\Ecm}$, where \Ecm\ 
      $= 2\times \Ebeam$.
\item Visible momentum along the beam direction: 
      $|\Sigma~p_{z}^{\mrm{vis}}|$.
\end{itemize}

\subsection{Photon Identification}
\label{s:photid}

Photon identification was performed using the high resolution lead glass 
electomagnetic calorimeter combined with information from the tracking
detectors. Photon candidates were classified as one of three types.
The first type was unassociated EC 
clusters; which were defined by the requirement that no charged track 
lie within the angular resolution of the cluster when extrapolated 
to the calorimeter. Furthermore, the lateral spread of the 
cluster was required to satisfy the criteria described in 
reference \cite{OPAL_HGG}.
The photon detection efficiency was increased
by considering two types of photon conversion candidates:
two-charged track conversions and ``single-track'' conversions.
Two-charged track conversions were selected as in reference \cite{2CTCONV}.
Effects such as finite two-track resolution and asymmetric 
conversions limit the ability to cleanly identify both electrons from
a converted photon. This inefficiency was addressed by considering
single-track conversions. 
A single-track conversion candidate required an EC cluster 
associated with a charged track consistent with pointing to
the primary vertex and having no associated hits in either layer
of the silicon micro-vertex detector or in the first 6 layers of
the central vertex chamber. Up to one additional charged track passing the
same criteria was allowed to point to the cluster. 
For both types of conversions, the photon energy was defined by the 
sum of cluster energies pointed to by the track(s).

Photon candidates were required to satisfy an isolation
requirement. The energy of additional tracks and clusters
in a $15\degree$ half-angle cone defined by the photon direction had to 
be less than 2 GeV. For each final state topology, the photon selection 
required:
\begin{itemize}
 \item At least two photon candidates in the fiducial region
       $15\degree < \theta < 165\degree$ 
       having $x_{\gamma}>0.05$ where $\xgam$ was 
       the photon energy scaled to the beam energy,
       $\xgam= E_{\gamma}/E_{\mrm beam}$ and $\theta$
       was the angle of the photon with respect to $\mrm e^-$ beam 
       direction.
 \item At least one of the photons was required to have $\xgam>0.10$.
\end{itemize}
For the case of more than two photon candidates in an event, 
only the two highest energy candidates were considered. 
The error on the di-photon mass was computed on an event-by-event
basis using the photon direction and energy information with
the energy resolution being the dominant source. For the mass range 
considered in this search, the di-photon invariant mass 
resolution (RMS)
can be parameterized as $\sigma_{\mdip} = 0.42$~GeV  + $0.02\mdip$.



\subsection{Hadronic Channel}
\label{s:qqgg}
The hadronic channel consisted of a $\gaga~+~hadrons$ final state. 
The hadronic channel was defined only by a charged track multiplicity 
requirement.
For the $\Hboson\Zboson$ topology, the dominant physics background
is the radiative return process $\Zboson\gaga$. Backgrounds also can arise
from radiative $\Zboson\gamma$ events where a decay product of the
\Zzero, {\em e.g.}, an isolated $\pi^0$ or $\eta$ meson, mimics a photon, 
or there is an energetic final state radiation (FSR) photon.
In these cases, the recoil mass against the di-photon system will tend to be 
lower than the \Zzero\ mass; therefore, this background can be suppressed by 
requiring a mass consistent with that of the \Zzero. 
In the general search topology, there is no {\em a priori}
mass constraint to help suppress backgrounds from fake photons.
Therefore, the backgrounds from radiative return events
are addressed by restricting the photon fiducial acceptance.


The hadronic channel candidate selection is summarized in Table
\ref{T:qq1}. Candidate events were 
required to satisfy the following criteria:
\begin{itemize}
  \item[(A1)] The standard hadronic event selection
              described in reference~\cite{hadsel}
              with the additional requirements:
    \begin{itemize}
       \item $\Rvis > 0.5$ and $|\Sigma~p_{z}^{\mrm{vis}}| < 0.6 \Ebeam$;
       \item at least 2 electromagnetic clusters 
             with $E/\Ebeam > 0.05$.
    \end{itemize}
   \item[(A2)] At least two photon candidates satisfying
               the photon selection criteria described in 
               Section \ref{s:photid}. In Table~\ref{T:qq1},
               the selection is broken down as (A2.A) and
               (A2.B), where (A2.A) is the requirement for  
               at least one photon with $\xgam>0.10$, and
               (A2.B) is the requirement of a second 
               photon with $\xgam>0.05$.

   \item[(A3)] To address the background from FSR and fake photons,
               the charged tracks and unassociated clusters were 
               forced into two jets using the Durham scheme \cite{Durham} 
               excluding the photon candidates. The lower energy photon 
               candidate was then required to satisfy  
               $p^{\gamma}_{\mrm jet} > 5$ GeV/$c$, where 
               $p^{\gamma}_{\mrm jet}$ was defined as the photon 
               momentum transverse to the axis defined by the closest jet. 

   \item[(A4.1)]For the $\Hboson\Zboson$ topology, it was required that 
                the invariant mass
                recoiling from the di-photon satisfy 
                $|M_{\mrm recoil} - M_{\mrm Z}| < 20$ GeV. 


   \item[(A4.2)] For the more general search topology, it was required that
                 the both photons satisfy $|\cos\theta_{\gamma}| <0.9$.
                 The recoil mass cut (A4.1) was not applied.

\end{itemize}

After applying the cut on the recoil mass, 14 events remained versus
an expectation based on the PYTHIA generator of $23.9\pm1.7$ events, 
where the error is statistical.\footnote{Unless otherwise specified, all errors
quoted will be statistical only.} For the high mass region, \mgg$>40$ GeV,
6 events were observed versus an expectation $15.2\pm1.4$.
For the general search topology, where no explicit
recoil mass cut was made, 13 events were observed versus $22.5\pm1.7$
expected after the cut on the photon polar angles. 

The contribution from $\pi^0$ and $\eta$ mesons misidentified as
photons from boosted \Zzero\ decays was 
estimated using data.  
Events having a photon satisfying $\xgam>0.6$, {\em i.e.} consistent 
with the hypothesis 
of being a photon from a radiative return to the \Zzero, were selected. 
Using charged tracks satisfying the same 
isolation criteria used in the photon selection as a proxy for isolated 
$\pi^0$ and $\eta$ mesons, and the measured $\pi^0,\eta$ and charged
track multiplicities \cite{PDG}, we estimated $5.5\pm1.7$ events from fakes 
before application of the $p^{\gamma}_{\mrm jet}$ 
criteria (A3) and $1\pm1$ afterwards.
No event passed the recoil mass cut.  
An identical study using Monte Carlo simulation
predicted $6.4\pm0.6$ events before 
the $p^{\gamma}_{\mrm jet}$ cut and $1.1\pm0.24$ afterwards. The recoil mass cut
reduced this to $0.50\pm0.16$ events expected from fake photons. 
Using the observed number of photons having $\xgam>0.1$ as a normalization,
we estimate the total number of events from fakes in the
\Hzero\Zzero\ and general search channels to be $0.8\pm0.2$ and $1.7\pm0.4$,
respectively.


The observed number of events for both searches is 
significantly smaller than the number expected from the PYTHIA simulation.
In contrast, the HERWIG generator predicts $8.6\pm1.0$ and 
$7.5\pm0.9$ events for the \Hzero\Zzero\ and general search channels, 
respectively, significantly lower than what is observed.
An analytic calculation predicts a cross section for $\Zboson\gaga$ of 
approximately 250 fb summed over all \Zzero\ decay modes~\cite{BBZgg}, 
corresponding to a cross section of 175 fb for hadronic decays of the 
\Zzero.
The calculation considered photons having transverse momentum greater than 
10 GeV/$c$ with respect 
to the beam axis and lying in the region $15\degree < \theta < 165\degree$.
The same criteria applied at the generator level 
to PYTHIA and HERWIG 
give cross sections of $261\pm24$ (stat.) fb and $115\pm15$ (stat.) fb, 
respectively.
If the expected number
of events from the PYTHIA simulation is rescaled by the ratio of
the generator level cross section to the analytic calculation, 
then the observed number of candidates, 14, is in agreement with
the expectation of $16.0\pm1.9$. Similarly, for the general search
topology, the rescaled expectation gives $15.1\pm1.9$ events which
accords with the 13 observed. Similar agreement is observed with 
HERWIG after the rescaling.
Given the uncertainty in the simulation
of the $\Zboson\gaga$ process, we do not perform a background 
subtraction when computing limits. 

The efficiency for this analysis to accept \Hzero\Zzero\ events 
for Higgs masses of 40 to 100 GeV is shown in Table~\ref{t:hzeff}.
For the general search, efficiencies were typically at least 50\% 
for most of the region for $10 < \MX < 170$ GeV with a degradation
to 30\% for $\MX<40$ GeV and $\MY$ near the kinematic limit.

\subsection{Charged Lepton Channel}
\label{s:llgg}

The exceptionally clean nature of the $\gaga \ellell $ final state
obviated requiring well-identified leptons. No distinction between 
the $\mrm e,\mu$ and $\tau$ channels was made. Leptons were defined 
as low multiplicity jets formed from charged tracks and isolated 
EC clusters.
By allowing single charged tracks to define a jet and with no explicit
lepton identification required, the selection maintained high efficiency
for tau leptons. The distinct di-photon signature was further exploited 
by including the single-charged track topology, $\gaga \ell (\ell)$,
where one of the leptons was missed due to tracking inefficiency 
near the beam axis.   
The most serious background was 
Bhabha scattering with initial and/or final state radiation. 
The large Bhabha scattering cross section necessitated a more restrictive 
fiducial region for accepting photons in both the \Hzero\Zzero\
and general searches. 



The leptonic channel event selection is summarized in Table~\ref{T:ggll}.
Leptonic channel candidates were required to satisfy the following
selection criteria:

\begin{itemize}
   \item[(B1)] Low multiplicity preselection~\cite{lowmsel} and:
 \begin{itemize}
   \item $\Rvis>0.2$ and $|\Sigma~p_{z}^{\mrm{vis}}|<0.8 \Ebeam$;
   \item number of EC clusters not associated with tracks:  
         $N_{\rm EC} \leq 10$;
   \item number of good charged tracks: 
         $1 \leq N_{\rm CT} \leq 7$;
   \item at least 2 electromagnetic clusters 
             with $E/\Ebeam > 0.05$.
  \end{itemize}
  \item[(B2)] At least two photon candidates satisfying the photon 
              selection described in Section \ref{s:photid}.


  \item[(B3)] For events having only one good charged track, require:
     \begin{itemize} 
       \item the track not be associated with a converted photon;
       \item the track have momentum satisfying $p>0.2E_{\mrm beam}$;
       \item direction of event missing momentum: 
             $|\cos\theta_{\mrm miss}| > 0.90$. 
     \end{itemize}

  \item[(B4)] For events having two or more good charged tracks, the event
              was forced to have 2 jets within the Durham scheme. Each
              jet was then required to have at least one charged track
              and $E_{\mrm jet}>3$ GeV. Tracks from identified photon 
              conversions were not included. 
              The jet finding was performed excluding the photon candidates.


  \item[(B5)] Both photons were required to satisfy $|\cos\theta_{\gamma}| <0.9$.

  \item[(B6.1)] For the \Hzero\Zzero\ search, the recoil mass to the
                di-photon was required to be consistent with the \Zzero:
              $|\Mrec - \MZ| < 20$ GeV.
\end{itemize} 
Prior to the cut on the photon polar angles (B5), there is poor agreement
between the data and Monte Carlo. This is attributed to the 
imperfect simulation of the Bhabha scattering background at small polar angles;
the number of expected events is sensitive to small deviations in the 
simulation given the large cross section. 
After application of the photon polar angle cut, the number of observed 
events 
was 9 versus an expectation of $11.1\pm0.8$ events from Standard Model 
sources. After the recoil mass requirement, the number of
observed events was 3, in good agreement with the expectation of $4.0\pm0.4$.
For the high mass di-photon region,
\mgg$ > 40$ GeV, 3 events were observed before application of the recoil mass
cut, consistent with the $4.7\pm0.5$ expected events. 
The efficiencies for 
$\Hboson\ra\gaga$ for Higgs masses of 40 to 100 GeV are given in 
Table \ref{t:hzeff}. The XY search maintained efficiencies of at least 
35\% for di-photon masses between 10 and 170 GeV for all $\MY$ considered in 
this analysis.





\subsection{Missing Energy Channel}
\label{s:nngg}

The missing energy channel was characterized by a pair of photons recoiling
against a massive, unobserved system. An irreducible Standard Model background
is the process $\epem \ra \nu\bar{\nu} \gaga$. 
Other potential backgrounds include 
$\epem \ra \gaga(\gamma)$ and radiative Bhabha scattering with
one or more unobserved electrons. 


The event selection for the missing energy channel is summarized
in Table~\ref{T:ggnn}.
Candidates in the missing energy channel were required to 
satisfy the following selection criteria:
\begin{itemize}
\item[(C1)] Low multiplicity preselection~\cite{lowmsel} with
            the further requirement that the event satisfy the    
            cosmic ray and beam-wall/beam-gas vetos
            described in reference \cite{photsel}, and:
    \begin{itemize}
     \item number of EC clusters not associated with tracks:  
         $N_{\rm EC} \leq 4$;
     \item number of good charged tracks: 
         $N_{\rm NCT} \leq 3$;
     \item $|\Sigma~p_{z}^{\mrm{vis}}|<0.8 \Ebeam$;
     \item at least 2 electromagnetic clusters 
               with $E/\Ebeam > 0.05$.
    \end{itemize}

\item[(C2)] At least two photon candidates satisfying the photon 
            selection described in Section \ref{s:photid}.
            Furthermore, the sum of the scaled photon energies 
            was required to satisfy $x_{\gamma1}+x_{\gamma2}~<~1.9$
            to suppress contributions from $\epem \ra \gaga$ and
            Bhabha scattering.

\item[(C3)] Consistency with the hypothesis that the di-photon system
            is recoiling from a massive body. 
      \begin{itemize}            
         \item The momentum component of the di-photon system in the plane 
               transverse to the beam axis: $p_T (\gaga)>0.05 E_{\mrm beam}$.
         \item Opening angle between the two photons in the plane 
               transverse to the beam axis: $\phi_{\gaga}<177.5\degree$.
         \item Polar angle of the di-photon system: 
               $|\cos\theta_{\gaga}| < 0.96$.
      \end{itemize}

\item[(C4)] Charged track veto: events were required to have no charged track
            candidates (other than those associated with an identified photon 
            conversion) as defined by the track veto criteria of reference 
            \cite{OPAL_HGG}. 


  \item[(C5)] Veto on unassociated calorimeter energy: the energy observed 
              in the  electromagnetic
            calorimeter not associated with the 2 photons was required to
            be less than 3 GeV.

  \item[(C6)] Both photons were required to satisfy 
              $|\cos\theta_{\gamma}| < 0.9$.

  \item[(C7.1)] For the \Hzero\Zzero\ search, the recoil mass against
                the di-photon was required to be consistent with the \Zzero:
              $|\Mrec - \MZ| < 20$ GeV.
\end{itemize}

After application of the photon fiducial cut (C6), two candidates
remain compared to an expectation of $4.7\pm0.1$ events from Standard Model
sources. The expected background at this
point is dominated by the irreducible $\epem \ra \nu\bar{\nu} \gaga$ process.
Subsequent application
of the recoil mass cut leaves two candidates versus $3.1\pm0.1$ expected.
No new physics processes are suggested. The efficiencies for Higgs masses
from 40 to 100  GeV for the \Hzero\Zzero\ search are summarized in 
Table \ref{t:hzeff}.
The efficiency for the general search channel was at least 50\% over
most of range of masses considered. 
For the extreme case of $\MX=10$ GeV and $\MY=160$ GeV, the
acceptance was 55\%; for $\MX=170$ GeV and $\MY=10$ GeV, 
the acceptance was 25\%, where the loss of efficiency was 
due to the requirement on the sum of the photon energies.

\section{Results}
\label{s:results}

\begin{figure}[!p]
        \vspace{0.8cm}
        \begin{center}
            \resizebox{\linewidth}{!}{\includegraphics{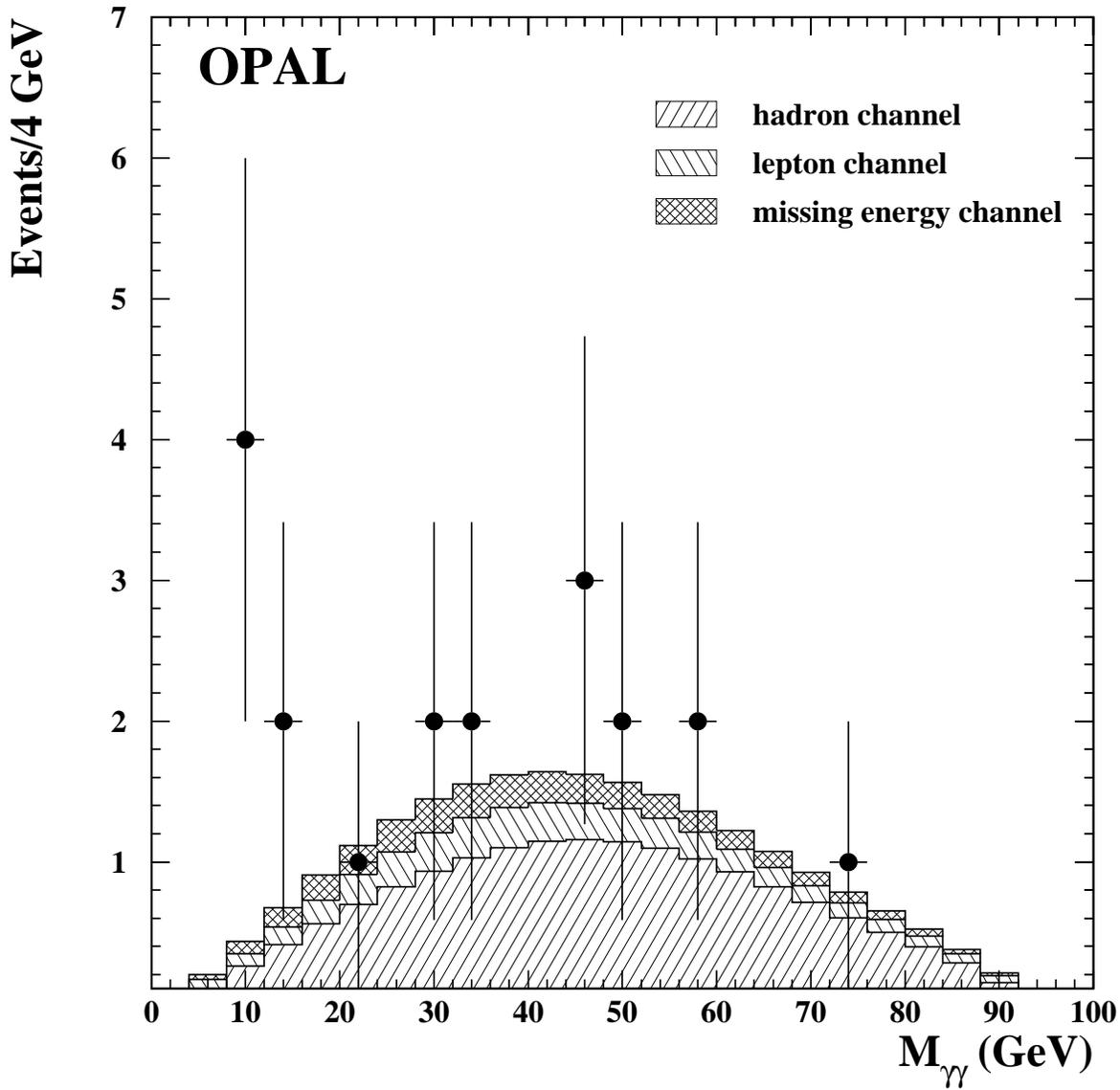} }
        \caption[COMGG]{    
                  Distribution of mass of the two highest energy photons
                  in the \Hzero\Zzero\ search 
                  after application of all selection criteria.
                  All search channels are included.
                  Data are shown as points with error bars.
                  Background simulation is shown as a histogram 
                  showing the contributions from the hadronic, 
                  charged lepton and missing energy channels as
                  denoted. The hadronic contribution has been
                  rescaled as described in Section \ref{s:qqgg}.
                  The background distributions have been
                  smoothed.
        \label{COMGG} }
        \end{center}
    \end{figure}

The di-photon invariant mass distribution for the events passing all cuts
in the \Hzero\Zzero\ topology is shown in Figure~\ref{COMGG};
the simulation of Standard Model backgrounds is also shown in the figure. 
For presentation purposes, the hadronic contribution has been rescaled 
as per the discussion in Section \ref{s:qqgg}. 
The highest di-photon mass candidate
consistent with recoiling from an associated \Zzero\
had a mass of $73.5\pm3.3$ GeV.
Summing over all \Zzero\ decay modes and expected background sources 
yields $23.1\pm1.9$ events expected versus 19 observed where 
the rescaled expectation in the hadronic channel has been used.

\begin{figure}[!p]
        \vspace{0.8cm}
        \begin{center}
            \resizebox{\linewidth}{!}{\includegraphics{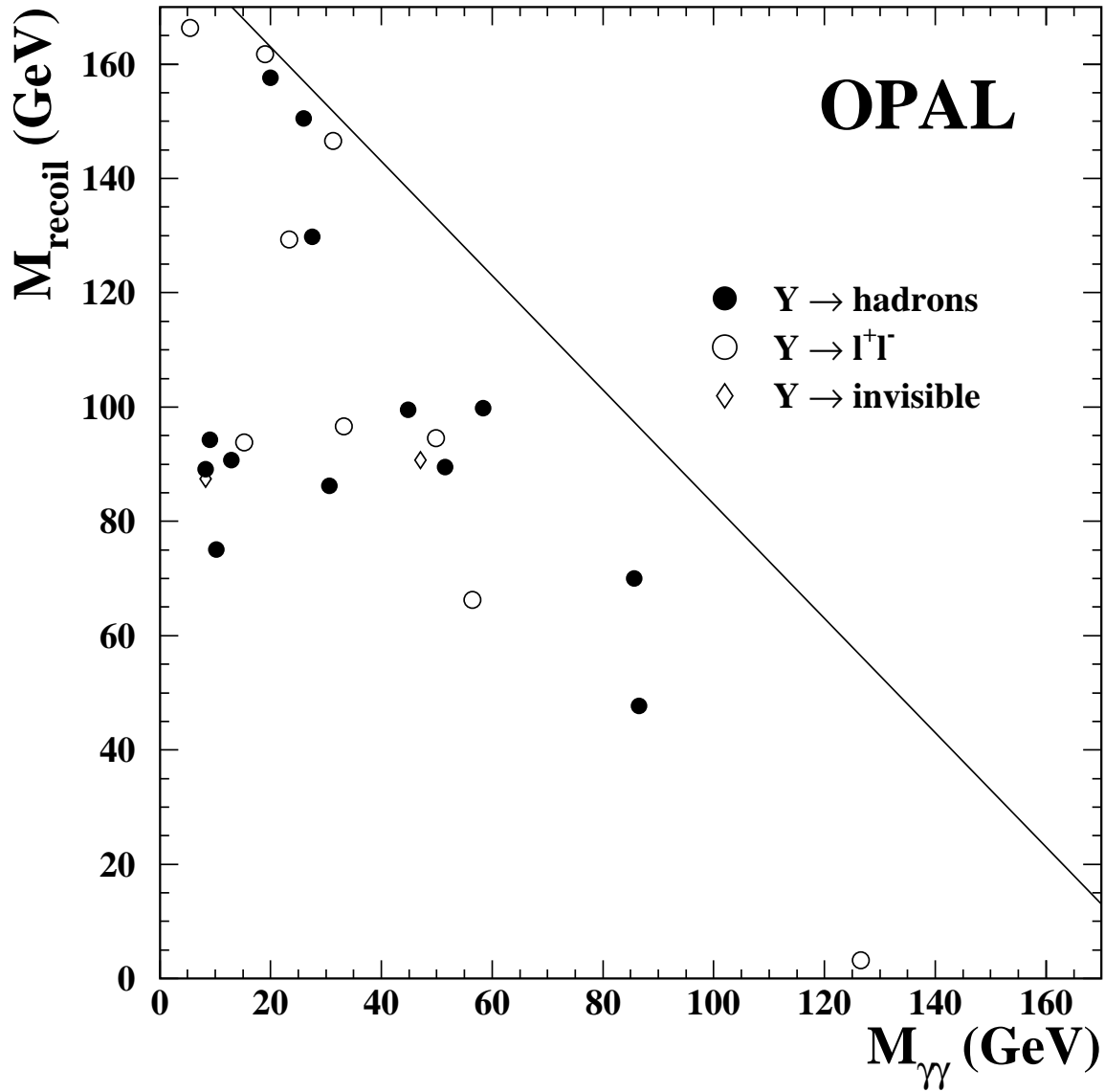} }
        \caption[COMGG2]{    
                  Distribution of mass recoiling against the di-photon system
                  versus di-photon invariant mass for the XY search.
                  The different search channels are as indicated. 
                  All selection criteria have been applied.
                  The diagonal line denotes the kinematic limit.
        \label{COMGG2} }
        \end{center}
    \end{figure}

For the general search, the recoil mass to the di-photon
is plotted versus the di-photon mass in Figure~\ref{COMGG2}.
Combining all search channels and using the rescaled hadronic
channel expectation results in
24 observed events versus $30.9\pm1.9$ expected from Standard Model sources. 
Two high di-photon mass
candidates were observed in the hadronic channel 
having masses of $85.6\pm2.0$~GeV and
$86.5\pm3.3$~GeV. Their respective recoil masses were $69.9$~GeV and
$47.7$~GeV with both events having their most energetic photon 
consistent with that from a 
radiative return to the \Zzero. The observation of two events of this
type is consistent with the $0.3\pm0.1$ events expected from 
a radiative return photon combined with a fake photon 
with the di-photon mass greater than 60 GeV. 




The systematic uncertainty in the photon detection efficiency,
3\% (relative) per photon, is primarily due to the simulation of the 
photon isolation criteria \cite{OPAL_ggjj_1}.
The contribution from the angular resolution and photon 
energy cuts was negligible. The energy scale for
high energy photons was determined to be reliable at the 0.6\% level.
This was inferred from the measured \Zzero\ mass of $90.58\pm0.40$ GeV 
using the mass recoiling against photons having $0.70 < \xgam < 0.82$.
The only other significant source was statistical error, 
typically 3.8\%, on signal samples used to determine the efficiency.
The systematic error on the integrated luminosity of the data, 0.46\% 
contributed negligibly to the limits. Variation of the non-photon
identification criteria led to effects smaller than the statistical
error in the acceptance. The efficiency for Higgs masses at intermediate
points was determined using a linear interpolation.
We conclude that the total systematic error on the 
acceptance in the \Hzero\Zzero\ search to be 6.0\% with a negligible 
dependence on the Higgs mass. 

From the events passing the cuts, the 95\% CL upper limit  
on the number of signal events
at a given di-photon mass was computed using the method of reference~\cite{BOCK}.
The error on the di-photon mass was computed for each event using the
measured errors of the photon positions and energies.
The effect of the systematic error on the upper limits was incorporated
using the method prescribed in reference \cite{CandH}.
Given the existing uncertainty in the simulation of the background
in the hadronic channel, no background subtraction was performed.



\begin{figure}[!htb]
        \vspace{0.8cm}
        \begin{center}
            \resizebox{\linewidth}{!}{\includegraphics{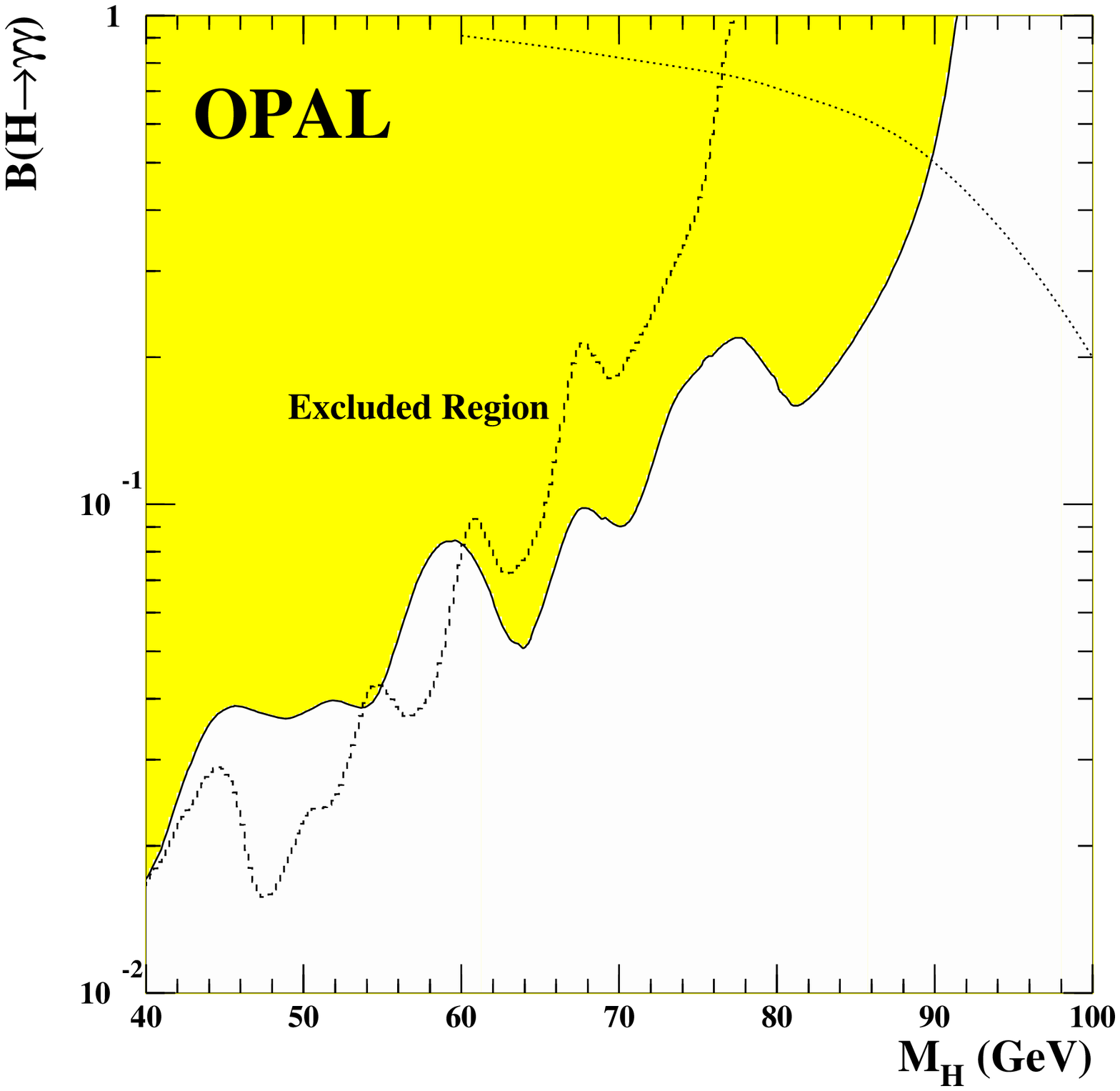} }
        \caption[bgglim]{
                 95\% confidence level upper limit on $B$($\Hboson \ra \gaga$)
                 for a Standard Model Higgs boson production rate. 
                 The shaded region is excluded. The dotted line
                 is the predicted $B$($\Hboson \ra \gaga$) in the limit of
                 $B$($\Hboson \ra \mrm f \bar{f}$)=0 \cite{Bosonic}. The
                 intersection with the exclusion curve gives a lower limit of
                 90.0 GeV for the Bosonic Higgs model. Candidates from
                 references \cite{OPAL_HGG} and \cite{OPAL_ggjj_1} have
                 been included. The dashed line indicates the previous
                 limit from reference \cite{OPAL_HGG}.
                 
        \label{bgglim} }
        \end{center}
    \end{figure}

In Figure~\ref{bgglim}, limits on $B(\Hboson \ra \gaga)$ 
up to $\MH = 92$ GeV are shown assuming a Standard Model \Hzero\Zzero\
production rate \cite{HZHA}. Included in the calculation of the
upper limit are the sensitivities and candidates from previous
OPAL searches at $\sqrt{s} \geq \MZ$ \cite{OPAL_HGG} and at 
the \Zzero\ \cite{OPAL_ggjj_1}.
For masses below 40 GeV, the inferred limits for $B(\Hboson \ra \gaga)$ from
references \cite{OPAL_ggjj_1,LowMgg} are more restrictive.
The limits on the $B(\Hboson \ra \gaga)$ 
can be used to rule out Higgs bosons in certain 
non-minimal models. 
In models with suppressed fermion couplings, the tree level 
decay mode to WW$^{(*)}$ is highly suppressed due to the off-shell W boson, 
resulting in a large di-photon branching ratio for 
$\MH < 100$ GeV. Figure \ref{bgglim} shows the $\Hboson \ra \gaga$ branching
ratio in the absence of fermionic decay modes as calculated in 
reference \cite{Bosonic}. 
In the Bosonic Higgs model \cite{Bosonic}, the Higgs 
scalar has Standard Model strength couplings to the $\mrm SU(2)$ gauge bosons but 
no tree-level couplings to fermions. From the extracted upper limits for 
$B( \Hboson \ra \gaga)$, Higgs scalars of this type are ruled out up
to a mass of $90.0$ GeV at 95\% CL.\footnote{Numerical mass limits are 
quoted to 0.5 GeV precision.}



\begin{figure}[!htb]
        \vspace{0.8cm}
        \begin{center}
            \resizebox{\linewidth}{!}{\includegraphics{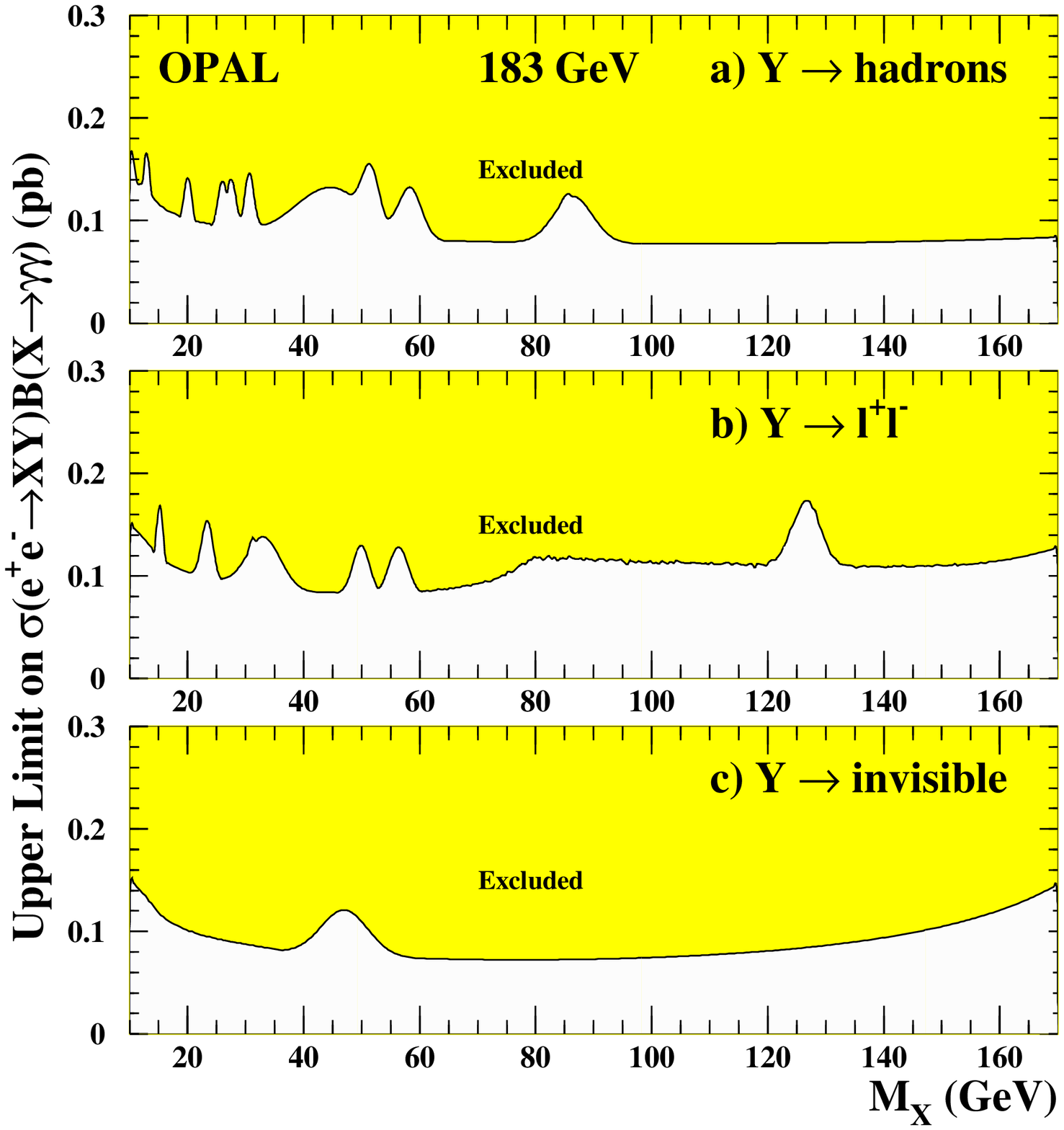} }
        \caption[limxy]{
                 95\% confidence level upper limit on
                 $\sigma(\epem\ra {\rm XY}) \times B(\mrm X \ra \gaga)$
                 for the case 
                 where: a) Y decays hadronically, b) Y decays into any charged
                 lepton pair and c) Y decays invisibly. The limits
                 for each $\MX$ assume the smallest efficiency as a function of
                 $\MY$ such that $10 < \MY < 160$ GeV and that
                 $\MX + \MY > M_{\mrm Z}$.
        \label{limxy} }
        \end{center}
    \end{figure}

For the general search channel, $\epem \ra \rm XY$ where X$\ra\gaga$ 
and Y$\ra \mrm f\bar{f}$, the results, in the form of upper 
limits on production cross section times di-photon branching fraction times
the branching ratio for Y to hadrons, charged leptons or invisible
are shown in Figure~\ref{limxy}. Only the 183 GeV data set has been
included. The limits are for 
scalar masses $10 < \MX < 170$ GeV such that 
$10 < \MY < 160$ GeV and $\MX + \MY > M_{\mrm Z}$.
The limits for each $\MX$ assume the weakest limit as a function of
$\MY$ subject to the above constraints.
The results assume that the production cross section has the same angular 
distribution as that of \Hzero\Zzero\ production. 
For a scalar/vector hypothesis for X and Y, the efficiency is
consistent at the 5\% level with that for a scalar/scalar hypothesis.
The systematic error in the efficiency from the interpolation near 
the kinematic edges is approximately 10\%; otherwise, it is consistent
with the statistical error in the number of accepted signal events.
For the Y$\ra \ell^+\ell^-$ channel, the
sensitivity was conservatively determined using the efficiency assuming 
a 100\% rate for Y$\ra \tau^+\tau^-$. The limits further assume that all
candidates in the charged lepton channel are consistent with a $\tau$-pair
hypothesis. Cross sections limits 
of 70 -- 200~fb are obtained over $10 < \MX < 170$ GeV with limits
of typically 100~fb for most values of $\MX$.

For the Higgs Triplet Model 
discussed in references \cite{Fermiophobic,Akeroyd},  
the results of the general search can be
used to set a mass limit on the neutral member of the five-plet, $\mrm H_5^0$.
We consider the production processes 
$\epem \ra \mrm H_5^0\Zboson$ and $\epem \ra \mrm H_5^0 H_3^0$ 
which, in a manner analogous to 2HDM models, have
their rates modified by complementary factors 
$4/3\sin^2\theta_H$ and $4/3\cos^2\theta_H$, 
respectively.\footnote{The parameter $\theta_H$ is defined as
$\sin^2\theta_H = 8b^2/(a^2 + 8b^2)$ where $a$ and $b$ are
the vacuum expectation values of the doublet and triplets,
respectively.}
In this model, the
five-plet is fermiophobic and the $\mrm H_3^0$ has similar properties
to the $\mrm A^0$ of 2HDM models. Members of the same multiplet will 
be degenerate in mass. The mass limit 
for charged Higgs bosons, 52 GeV at the 95\% CL \cite{ChargedH} applies to the
charged member of the three-plet and phenomenological
constraints require the members of the five-plet to be more massive than the three-plet
\cite{Akeroyd}. For each value of $M_{\mrm H_5^0}$, the minimum number
of expected events as a function of $\sin^2\theta_H$ and $M_{\mrm H_3^0}$
was computed and compared to the least restrictive general search channel
result at that mass.
If it is assumed that decays between the multiplets can be neglected, 
{\em e.g.} $\mrm H_5^0 \ra H_3^0 Z^{(*)}$, then a 95\% CL lower limit
of 79.5 GeV for the mass of the $\mrm H_5^0$ is obtained. 
This limit assumes that $\mrm H_3^0$ decays exclusively to 
fermion pairs. 


\section{Conclusions}

Using a data sample of 
57.7~\ipb\ taken at a luminosity weighted center-of-mass energy of 
182.6 GeV, a  search for the production of Higgs bosons and other
new particles 
decaying in photons has been performed. No evidence of resonant 
behavior in the di-photon mass spectrum was observed.
The results of this search have been
combined with previous OPAL results to set limits on $B$($\Hboson \ra \gaga$) 
up to a Higgs boson mass of 92 GeV, provided the Higgs particle is produced 
via $\epem \ra \Hboson \Zboson$ at the Standard Model rate.
A lower mass bound of 90.0 GeV is set at the 95\% confidence level for
Higgs particles which couple only to gauge bosons but still couple to
the \Zzero\ at Standard Model strength. 
Model independent upper limits on
$\sigma(\epem\ra {\mrm X Y})\times B({\mrm X} \ra \gaga)\times 
B(\mrm Y \ra f\bar{f})$
of 70 -- 200~fb are obtained over $10 < \MX < 170$ GeV where
$10 < \MY < 160$ GeV and $\MX + \MY > M_{\mrm Z}$ independent of
the scalar/vector nature of Y provided that the Y decays to a fermion pair
and has negligible width.
For Higgs Triplet models having custodial $\mrm SU(2)$ symmetry,
these results rule out masses up to 79.5 GeV at the 95\% CL 
for members of the five-plet
provided that decays involving lower mass Higgs multiplets can be neglected.
%
%
\bigskip\bigskip\bigskip
\begin{flushleft}
{\Large\bf Acknowledgements}
\end{flushleft}
\par
We particularly wish to thank the SL Division for the efficient operation
of the LEP accelerator at all energies
 and for their continuing close cooperation with
our experimental group.  We thank our colleagues from CEA, DAPNIA/SPP,
CE-Saclay for their efforts over the years on the time-of-flight and trigger
systems which we continue to use.  In addition to the support staff at our own
institutions we are pleased to acknowledge the  \\
Department of Energy, USA, \\
National Science Foundation, USA, \\
Particle Physics and Astronomy Research Council, UK, \\
Natural Sciences and Engineering Research Council, Canada, \\
Israel Science Foundation, administered by the Israel
Academy of Science and Humanities, \\
Minerva Gesellschaft, \\
Benoziyo Center for High Energy Physics,\\
Japanese Ministry of Education, Science and Culture (the
Monbusho) and a grant under the Monbusho International
Science Research Program,\\
German Israeli Bi-national Science Foundation (GIF), \\
Bundesministerium f\"ur Bildung, Wissenschaft,
Forschung und Technologie, Germany, \\
National Research Council of Canada, \\
Research Corporation, USA,\\
Hungarian Foundation for Scientific Research, OTKA T-016660, 
T023793 and OTKA F-023259.\\

\bigskip\bigskip\bigskip\bigskip\bigskip\bigskip

\newpage


\def\mulc1{\multicolumn{1}{|c|}}

\begin{table}[htbp]
\begin{center}
\begin{tabular}{|l||r||r|r|r|}\hline
 Cut  & Data & \mulc1{$\Sigma$Bkgd} &\mulc1{$(\gamma/{\rm Z})^{\ast}$} & \mulc1{4f} \\ 
\hline\hline
 (A1)    &  3303 &  3421   &   2620    &    801.4  \\ \hline
 (A2.A)  &    672  &   765.3 &    732.1  &     33.2  \\ \hline
 (A2.B)  &     38  &    46.3 &     45.2  &      1.1  \\ \hline
 (A3)    &     24  &   35.6 &     35.0  &      0.6   \\ \hline
\hline
 (A4.1)  &   14  &     $23.9\pm1.7$ &     $23.7\pm1.7$  &  $0.2\pm0.1$ \\ \hline
\hline
 (A4.2) &     13  &     $22.5\pm1.7$ &     $22.2\pm1.7$  & $0.3\pm0.1$  \\ \hline 
\end{tabular}
\end{center}
  \caption[Data and MC after cuts]
  {Events remaining in the hadronic channel search after the indicated cumulative cuts.
   The selection criteria are as described in Section \ref{s:qqgg}. 
   The entry for 
   (A4.1) is for the $\Mrec$ cut for the \Hzero\Zzero\ search. The 
   entry for (A4.2) corresponds to the final photon fiducial cut employed
   in the general search channel. Note that the $\Mrec$ cut has not been
   applied for the general search channel. 
   In addition to the total simulated background, the simulations for
   $(\gamma/{\rm Z})^{\ast}$, four-fermion (``4f") final states
   are shown. For $(\gamma/{\rm Z})^{\ast}$, the unscaled prediction
   from Pythia is shown (see Section \ref{s:qqgg}).
   The background simulation samples are scaled to 57.7~\ipb.}
  \label{T:qq1}

\end{table}


\begin{table}[!htbp]                      
\begin{center}
\begin{tabular}{|l||r||r|r|r|r|r|r|}\hline
 Cut     & Data   &\mulc1{$\Sigma$Bkgd} &\mulc1{$\epem$} & \mulc1{$\tptm$}  
                                        & \mulc1{$\mpmm$} & \mulc1{$\gaga$}  
                                                             & \mulc1{\eeff}  \\ 
\hline\hline
 (B1)  & 13988 &   2935 &   2548 &  61.6 &    8.8 &   92.2 &   224   \\ \hline
 (B2)  &   405 &    276 &    202 &   6.4 &    3.3 &   62.3 &     2.3 \\ \hline
 (B3)  &   236 &    156 &    112 &   5.3 &    3.1 &   34.1 &     1.6 \\ \hline
 (B4)    & 49  &  35.5   &  28.3   &  3.2    & 3.0    & 0.5   &  0.5   \\ \hline

 (B5)    &  9  & $11.1\pm0.8$ & $7.3\pm0.8$ &  $1.7\pm0.1$    & $1.9\pm0.1$    
                                               & $0.1\pm0.1$   &  0.0   \\ \hline \hline
 (B6.1)  &  3  & $4.0\pm0.4$  & $2.2\pm0.4$   & $0.8\pm0.1$    
                                               & $0.9\pm0.1$ & 0.0 &  0.0   \\ \hline
\end{tabular}
\end{center}
  \caption[Data and MC after cuts]
  {Events remaining for leptonic channel analysis after the indicated 
   cumulative cuts.
   The selection criteria are as described in Section \ref{s:llgg}. 
   The contributions from Bhabha scattering ($\epem$),
   $\mu$-pair, $\tau$-pair production, $\gaga$ and \eeff\ final states 
   determined from background simulations are shown. The simulated datasets 
   have been normalized to 57.7 \ipb. Criteria (B6.1), the recoil mass cut,
   is only applied for the \Hzero\Zzero\ search. The poor agreement prior
   to the photon fiducial cut (B5) is due to imperfect simulation of
   Bhabha scattering at small polar angles.
}  \label{T:ggll}
\end{table}

%

\begin{table}[htbp]
\begin{center}
\begin{tabular}{|l||r||r|r|r|r|r|r|}\hline
 Cut  & Data & \mulc1{$\Sigma$Bkgd} & \mulc1{$\nunu\gaga$} & \mulc1{$\gaga$} 
                     & \mulc1{$\epem$} & \mulc1{$\ell^+\ell^-$} & \mulc1{\eeff}  \\ \hline\hline
 (C1)   & 66907 & 15925 &  14.5 & 730 & 14737 &  13.0 &  429   \\ \hline
 (C2)   &  1462 &  1456 &   9.4 & 140 &  1296 &   1.7 &    9.0 \\ \hline
 (C3)   &   139 &   125 &   8.5 &15.1 &  99.7 &   1.1 &    0.7 \\ \hline
 (C4)   &    28 &  26.1 &   8.1 & 13.2 & 4.7  &   0.1 &    0.0 \\ \hline
 (C5)   &     9 &   9.6 &   7.9 &  0.2 & 1.5  &   0.0 &    0.0 \\ \hline

 (C6)   &     2 & $4.7\pm0.1$ & $4.5\pm0.1$ & $0.0$ & $0.1\pm0.1$  & 0.0 & 0.0 \\ \hline \hline

 (C7.1) &     2 & $3.1\pm0.1$ & $3.1\pm0.1$ &  0.0 & 0.0  &   0.0 &    0.0 \\ \hline
\end{tabular}
\end{center}

  \caption[Data and MC after cuts]
  {Events remaining after the indicated cumulative cuts for the 
   missing energy channel search. 
   The selection criteria correspond to those described in Section \ref{s:nngg}. 
   The contributions from 
   $\nunu\gaga$, $\gaga$, $\epem$-pair, lepton pair ($\ell\equiv\mu,\tau$) 
   production and \eeff\ final states
   determined from background simulations are shown. The simulation datasets
   have been normalized to 57.7 \ipb. Criteria (C7.1), the recoil mass cut,
   is only applied for the \Hzero\Zzero\ search.

}  \label{T:ggnn}
\end{table}


\begin{table}[htbp]
\begin{center}
 \begin{tabular}{|l||c|c|c|c|c|c|c|c|}
\cline{2-9}
\multicolumn{1}{c}{ }  & \multicolumn{8}{|c|}{Higgs Mass (GeV)} \\ \hline 
 Channel       & 40   &   50  &    60  &    70  &    80  &    90  & 95 &   100 \\
\hline\hline
$\qqbar\gaga$   &  62 &  70 &  69 &  73 &  69 &  66 & 52 & 34 \\ \hline
$\ell\ell\gaga$ & 53 & 57 & 61 & 61 & 62 & 60 & 48 & 30 \\ \hline
$\nn\gaga$      &  60 & 61 & 65 & 67 & 70 & 63 & 50 & 32  \\ \hline

\end{tabular}
\end{center}
\caption[Acceptances for \Hzero\Zzero\ search channel.]
{Efficiency in percent (\%) for each \Hzero\Zzero\ search channel for Higgs masses
as indicated. For each \Zzero\ decay mode, 1000 events were generated.}
\label{t:hzeff}
\end{table}


\newpage
\newpage

\newpage

\newpage

\newpage


\begin{thebibliography}{99}

\bibitem{HBR} 
      J. Ellis, M. K. Gaillard, and D. V. Nanopoulos, \NPhys\ {\bf B106} (1976) 292; \\
      A. Djouadi, J. Kalinowski and M. Spira, \CPC{108} (1998) 56.     


\bibitem{Hagiwara} K.~Hagiwara and M.L.~Stong, \ZPhysC{62} (1994) 99.




\bibitem{Fermiophobic} 
      A. G. Akeroyd, \PhysLettB{368} (1996) 89.

\bibitem{Bosonic}
     A. Stange, W. Marciano, and S. Willenbrock, \PhysRev\ {\bf D49} (1994) 1354.

\bibitem{TypeI} 
     H. Haber, G. Kane and T. Sterling, Nucl. Phys. {\bf B161} (1979) 493.

\bibitem{Triplets} H. Georgi and M. Machacek, \NPhys\ {\bf B262} (1985) 463.


\bibitem{Gunion} J. F. Gunion, R. Vega and J. Wudka, 
                 \PhysRev\ {\bf D42} (1990) 1673.
 

\bibitem{OPAL_HGG} 
          \OPALColl, K.~Ackerstaff \etal, \EurPhysC{1} (1998) 31.

\bibitem{OPAL_ggjj_1}
     \OPALColl, G.~Alexander \etal, \ZPhysC{71} (1996) 1.

\bibitem{LowMgg}
     L3 Collab., M.~Acciarri \etal,  \PhysLett\ {\bf 388} (1996) 409; \\
     DELPHI Collab., P.~Abreu \etal, \ZPhysC{72} (1996) 179.



\bibitem{other_gg}
     \OPALColl, P.~Acton \etal, \PhysLett\ {\bf 311} (1993) 391; \\
     ALEPH Collab., D. Buskulic \etal, \PhysLettB{313} (1993) 299; \\
     L3 Collab., O.~Adriani \etal,  \PhysLett\ {\bf 295} (1992) 337.



\bibitem{detector} 
     \OPALColl, K.~Ahmet \etal, \NIMA{305} (1991) 275; \\
     O.Biebel \etal,      \NIMA{323} (1992) 169;         \\
     M.Hauschild \etal,   \NIMA{314} (1992)  74;  \\
      S.~Anderson \etal, \NIMA{403} (1998) 326.        

\bibitem{HZHA} HZHA generator: P.~Janot, in {\em Physics at LEP2,}
               edited by G.~Altarelli, T.~Sj\"{o}strand and
               F.~Zwirner, CERN 96-01 (1996) Vol.~2  p.309.

\bibitem{PYTHIA} 
     PYTHIA 5.721 and JETSET 7.408 generators: \\
     T. Sj\"{o}strand, \CPC{82}
     (1994) 74; \\
     T.\ Sj\"{o}strand, LUTP 95-20; \\
     ``PYTHIA 5.7 and JETSET 7.4, Physics and Manual", CERN-TH. 7112/93. \\
     The sample was generated with {\tt MSTP(68)=2} option for
     initial state radiation.

\bibitem{jtparams} 
     \OPALColl, G.~Alexander \etal, \ZPhysC{69} (1996) 543.

\bibitem{HERWIG} G.~Marchesini et~al., \CPC{67} (1992) 465.

\bibitem{BHWIDE}
     S.~Jadach, W.~Placzek and B.~F.~L.~Ward, University of Tennessee
     preprint, UTHEP~95-1001 (unpublished).

\bibitem{TEEGG}
     D. Karlen, \NPhys\ {\bf B289} (1987) 23.

\bibitem{KORALZ}
     S.~Jadach et al., \CPC{66} (1991) 276.

\bibitem{RADCOR} 
     F.A.~Berends and R.~Kleiss, \NPhys\ {\bf B186} (1981) 22.

\bibitem{VERMASEREN}
     J.~A.~M.~Vermaseren, Nucl. Phys. {\bf B229} (1983) 347.

\bibitem{grc4f} 
     J.~Fujimoto \etal, {\em GRC4F V1.1: A four fermion event generator for
     $\epem$ collisions}, preprint KEK-CP-046 and e-Print Archive: hep-ph/9605312.

\bibitem{excalibur}
     F.A.\ Berends, R.\ Pittau and R.\ Kleiss, \CPC{85} (1994) 43.


\bibitem{KORALW} 
 M.\ Skrzypek \etal, \CPC{94} (1996) 216;\newline
 M.\ Skrzypek \etal, \PhysLettB{372} (1996) 289. 




\bibitem{GOPAL} 
  J.Allison \etal, \NIMA{305} (1992) 47.

\bibitem{CTSEL}
     \OPALColl, G.~Alexander \etal, \ZPhysC{72} (1996) 191.

\bibitem{2CTCONV}
     \OPALColl, P.~Acton \etal, \ZPhysC{55} (1992) 191.


\bibitem{hadsel}
     \OPALColl, G.~Alexander \etal, \ZPhysC{52} (1991) 175.


\bibitem{Durham}
     N.~Brown and W.J.~Stirling, Phys.\ Lett.\ {\bf B252} (1990) 657; \\
     S.\ Bethke, Z.\ Kunszt, D.\ Soper and W.J.\ Stirling,
                                    Nucl.\ Phys.\ {\bf B370} (1992) 310; \\
     S.\ Catani \etal, Phys.\ Lett.\ {\bf B269} (1991) 432; \\
     N.\ Brown and W.J.\ Stirling, Z.\ Phys.\ {\bf C53} (1992) 629.

\bibitem{PDG} Particle Data Group, {\it Review of Particle Properties},
              \PhysRev\ {\bf D54} (1996) 1.

\bibitem{BBZgg} {\em Physics at LEP2}, Eds. G. Altarelli, T. Sj\"{o}strand, 
                and F. Zwirner, 
                CERN 96-01 (1996) Vol.~1 \\ pg. 244, as calculated in 
                G. B\'elanger and F. Boudjema, \PhysLett\ {\bf 288} (1992) 201.

\bibitem{lowmsel}
     \OPALColl, R.~Akers \etal, \ZPhysC{61} (1994) 19.

\bibitem{photsel}
     \OPALColl, R.~Akers \etal, \ZPhysC{65} (1995) 47.

\bibitem{BOCK}
     P. Bock, {\em Determination of exclusion limits for particle
     production using different decay channels with different efficiencies,
     mass resolutions and backgrounds}, 
     Heidelberg University preprint HD-PY-96/05 (1996);
     (submitted to \NIM).

\bibitem{CandH} 
     R.~D.~Cousins and V.~L.~Highland, \NIMA{320} (1992) 331.

\bibitem{Akeroyd} 
      A. G. Akeroyd, \PhysLettB{353} (1995) 519.

\bibitem{ChargedH} 
     \OPALColl, K.~Ackerstaff \etal, CERN-PPE/97-168, to be published in \PhysLett\ B.






\end{thebibliography}
\end{document}